\documentclass[letter]{aa}
\usepackage{graphicx, graphics}
\usepackage{amsmath, amssymb, bm}
\newcommand{\Ang}{\ensuremath{\text{\AA}}}
\usepackage[utf8]{inputenc}
\usepackage[hidelinks]{hyperref}
\usepackage{txfonts}
\usepackage[switch]{lineno}
\usepackage{booktabs}
\usepackage{multirow}
\usepackage{orcidlink}
\usepackage{placeins}
\titlerunning{Optical spectroscopy of WISPIT~2}
\authorrunning{Swastik et al.}
\begin{document}
\title{A Quiet Host in an Active Planet-Forming Disk: Optical Spectroscopy of WISPIT 2}
\author{
C. Swastik\orcidlink{0000-0003-1371-8890}\inst{\ref{unimi}}
\and M. Bestha\orcidlink{0009-0002-3354-3549}\inst{\ref{inst-iia}}\fnmsep\thanks{M. Bestha and L.~S. Sonith contributed equally to this work.}
\and L.~S. Sonith\orcidlink{0000-0002-2033-3051}\inst{\ref{inst-iia}}\fnmsep\footnotemark[1]
\and S. Facchini\orcidlink{0000-0003-4689-2684}\inst{\ref{unimi}}
\and T. Sivarani\orcidlink{0000-0003-0891-8994}\inst{\ref{inst-iia}}
\and R.~K. Banyal\orcidlink{0000-0003-0799-969X}\inst{\ref{inst-iia}}
\and Z. Wahhaj\orcidlink{0000-0001-8269-324X}\inst{\ref{inst-eso-cl}}
\and\\ A. Choudhary\orcidlink{0009-0004-6657-2260}\inst{\ref{inst-iitism}}
\and M. Gopinathan\orcidlink{0000-0002-1369-0608}\inst{\ref{inst-iia}}
\and P. Saraf\orcidlink{0009-0001-4813-0432}\inst{\ref{inst-prl}}
\and A. Mallick\orcidlink{0000-0002-4282-605X}\inst{\ref{inst-tifr}}
\and M.~P. Navaneeth\orcidlink{0009-0001-9837-5390}\inst{\ref{inst-iia}}
\and S. Biswas\orcidlink{0009-0000-2401-9986}\inst{\ref{inst-iia}}
\and K. Khushbu\orcidlink{0009-0004-8582-1373}\inst{\ref{inst-iia}}
}
\institute{
Dipartimento di Fisica, Universit\`a degli Studi di Milano, Via Celoria 16, 20133 Milano, Italy;\\ \email{swastik.chowbay@unimi.it} \label{unimi}
\and
Indian Institute of Astrophysics, II Block, Koramangala, Bengaluru 560034, India \label{inst-iia}
\and
European Southern Observatory, Alonso de C\'ordova 3107, Vitacura Casilla 19001, Santiago, Chile \label{inst-eso-cl}
\and
Physical Research Laboratory, Navrangpura, Ahmedabad 380009, India \label{inst-prl}
\and
Tata Institute of Fundamental Research, Homi Bhabha Road, Colaba, Mumbai 400005, India \label{inst-tifr}
\and
Indian Institute of Technology (Indian School of Mines) Dhanbad, Dhanbad 826004, Jharkhand, India \label{inst-iitism}
}
\date{Accepted 15 July 2026}
\abstract
{WISPIT~2 is a young pre-main-sequence star hosting a multi-ringed transition disk and two directly imaged protoplanets. One of them, WISPIT~2b, is confirmed to be accreting. This makes WISPIT~2 the closest known analogue to PDS~70; however, prior to this work, no optical spectrum of the central star had been published. Population-synthesis models predict that stellar accretion can be strongly suppressed once two or more giant planets have formed, so characterising the atmospheric and accretion properties of such hosts is essential.}
{We present the first optical spectrum of WISPIT~2 and use it to derive the host atmospheric parameters, test its youth spectroscopically, and constrain its stellar accretion state relative to PDS~70 and to the embedded planets.}
{We analysed low-resolution HFOSC spectra ($R\sim1200$ for Grism~7 and $R\sim2200$ for Grism~8) from the 2-m Himalayan Chandra Telescope with \texttt{iSpec}, validating the pipeline at HFOSC resolution against Gaia FGK Benchmark Stars and K-type pre-main-sequence templates from the public Manara et al. library. The H$\alpha$ residual filling was measured relative to synthetic photospheric profiles and compared with the expected chromospheric-noise level.}
{H$\alpha$ remains in net absorption but is partially filled in by emission: the weak residual filling ($1.5$--$2.0\sigma$) lies $\sim$1~dex below the chromospheric-noise level, consistent with chromospheric emission and no detectable accretion. We place a 95\% upper limit $\dot{M}_{\rm acc}<3.6\times10^{-11}\,M_\odot\,{\rm yr}^{-1}$, below even the lowest epoch of PDS~70's monitoring and implying a host-to-planet ratio $\dot{M}_\star/\dot{M}_{\rm 2b}\lesssim18$. We also measure $T_{\rm eff}=4551\pm150$~K and a first, low-resolution, model-dependent global metallicity $\mathrm{[M/H]}=-0.17\pm0.16$; the surface gravity and Li\,{\sc i} EW support the pre-main-sequence nature of the host.}
{Both known double-protoplanet hosts, PDS~70 and WISPIT~2, therefore appear to show strongly suppressed or undetectable stellar accretion. Larger spectroscopic samples are needed to test whether this is common in multi-protoplanet transition disks.}
\maketitle
\nolinenumbers
\section{Introduction}
\label{sec:intro}
Direct imaging of an accreting protoplanet embedded in its natal disk is exceptionally rare, requiring a young gas-rich transition disk, a planet accreting strongly enough to outshine the disk in H$\alpha$, and a favourable geometry. Only a handful of candidates have been confirmed by direct H$\alpha$ detection \citep{Haffert2019,Close2025}, and only two systems host more than one directly imaged protoplanet: PDS~70 \citep{Keppler2018,Haffert2019} and, since 2025, WISPIT~2 \citep{vanCapelleveen2025,Close2025,Lawlor2026}, making WISPIT~2 the closest known analogue to PDS~70. VLT/SPHERE imaging revealed an extended multi-ringed disk around a $\sim$5~Myr, $\sim$1.08\,$M_\odot$ pre-main-sequence star with an accreting $\sim$5\,$M_{\rm Jup}$ protoplanet (WISPIT~2b) at $\sim$57~au \citep{vanCapelleveen2025}; MagAO-X confirmed its H$\alpha$ accretion \citep{Close2025}; VLTI/GRAVITY revealed a second embedded companion (WISPIT~2c, $\sim$8--12\,$M_{\rm Jup}$ at $\sim$15~au) \citep{Lawlor2026}; and ALMA Band~7 resolved a narrow mm dust ring at 144~au, interior to which WISPIT~2b orbits \citep{Facchini2026}.
The discovery of two giant protoplanets in one disk makes it possible to test how accretion is divided between the star and the planets. In single-protoplanet systems gas is expected to leak past the gap-opening planet and accrete onto the star near the canonical T~Tauri value, but whether this persists with multiple giant planets is unclear. PDS~70 provides one data point, its host accreting more than an order of magnitude below the canonical rate \citep{Thanathibodee2020,Manara2023}; whether this suppression is generic to multi-protoplanet hosts or specific to PDS~70 cannot be settled without a comparable measurement on a second system.
The central star of WISPIT~2 remains poorly characterised, with the parameters of \citet{vanCapelleveen2025} deriving from Gaia astrometry and photometry and no published optical spectrum. In this Letter we present the first optical spectrum of WISPIT~2, obtained with HFOSC\footnote{Himalayan Faint Object Spectrograph Camera; \url{https://www.iiap.res.in/centers/iao/facilities/hct/}} on the 2-m Himalayan Chandra Telescope, and use it to derive the host atmospheric parameters and a low-resolution global [M/H] in the same \texttt{iSpec} framework applied to PDS~70 \citep[][PDS~70: $\mathrm{[Fe/H]} = -0.11 \pm 0.01$]{Swastik2021}, to test the system's youth from lithium, and to constrain the host accretion state from H$\alpha$. Given the modest resolution and PMS nature, we validate the methodology against Gaia FGK Benchmark Stars and K-type PMS empirical templates \citep{Manara2017}, both degraded to HFOSC resolution. We detect no stellar accretion and place an upper limit a factor of a few below PDS~70's central rate and well below the canonical T~Tauri rate, so both known double-protoplanet hosts appear to show strongly suppressed or undetectable host accretion.
\section{Observations and data reduction}
\label{sec:obs}
WISPIT~2\footnote{WISPIT~2 is the likely optical counterpart of the RASS X-ray source RX~J1923.2$-$0740 \citep[][their Table~2]{Moran1996}, providing an independent indicator of youth, consistent with the Li-rich, K-type RASS-selected young stars found around nearby star-forming regions \citep{Alcala1995,Krautter1997}.} (RA $=\,$19$^{\rm h}\,$23$^{\rm m}\,$17.03$^{\rm s}$, Dec $=\,-$07$\degr\,$40$'\,$55.1$''$, J2000; $V=11.60\pm0.12$~mag) was observed on UT 2026-04-14 with the HFOSC spectrograph on the 2-m Himalayan Chandra Telescope at Hanle, under a Director's Discretionary Time allocation, using the 1.92$''$-wide long slit (wider than the typical Hanle seeing, to minimise wavelength-dependent slit losses). One 600-s Grism~7 exposure ($\sim$3800--7700~\Ang, $R\sim1200$) was followed by three Grism~8 exposures ($\sim$5200--9300~\Ang, $R\sim2200$; 600, 600, 300~s), with Feige~34 observed the same night for flux calibration.

The data were reduced with standard \texttt{IRAF} routines (bias subtraction, flat-fielding, \texttt{lacos\_spec} cosmic-ray rejection, optimal extraction), wavelength-calibrated against FeAr/FeNe arcs (rms $\sim$0.1~pixel) and response-corrected using Feige~34. The resulting Grism~7 and Grism~8 spectra were combined and scaled to match the ASAS-SN $g$-band magnitude obtained from the light curve \citep{Shappee2014,Kochanek2017}, thereby placing them on an absolute flux scale.
The three Grism~8 frames are the primary dataset for all measurements: they resolve the narrow H$\alpha$ and Li\,{\sc i}~6708~\Ang\ diagnostics, cover the Ca\,{\sc ii} infrared triplet, and provide a frame-to-frame scatter estimate; the Grism~7 frame serves only as a broadband-continuum check (Fig.~\ref{fig:full_spec}).
Each spectrum was placed on the stellar rest frame by cross-correlating against a Castelli/ATLAS9 K-dwarf template ($T_{\rm eff}\simeq4550$~K, $\log g\simeq4.3$, [M/H]$=0$) broadened to HFOSC resolution, removing the residual $\sim$0.1--0.2~pixel arc zero-point; this matters for the absolute H$\alpha$ EW (Sect.~\ref{sec:halpha}) but is negligible for the broadband fits. Spectra were continuum-normalised by a cubic-spline fit over $\sim$50 line-free windows and median-combined within each grism. The combined Grism~8 spectrum reaches a continuum S/N per resolution element of $\sim$350--370 across 5700--8000~\Ang; telluric bands were masked.

\section{Results}
\label{sec:results}
\subsection{Optical spectrum}
\label{sec:spectrum}
The combined, continuum-normalised HFOSC spectrum (Fig.~\ref{fig:full_spec}, Appendix~\ref{app:posterior}) is that of a late-type K dwarf: the G-band, the Mg\,{\sc i}~b triplet, the Na\,{\sc i}~D doublet, Li\,{\sc i}~6708~\Ang, and the Ca\,{\sc ii} infrared triplet are all detected in absorption. The Ca\,{\sc ii}~IRT cores appear somewhat shallower than typical for a quiescent K3--K4V dwarf, tentatively suggesting weak chromospheric core infilling; we do not use the IRT cores as an accretion diagnostic at this resolution.
\subsection{Atmospheric parameters}
\label{sec:ispec}
We derived the atmospheric parameters with \texttt{iSpec} \citep{BlancoCuaresma2014a}, following \citet{Swastik2021}, using only the median-combined Grism~8 spectrum. This grism provides the higher-resolution red-optical range used for the fit; the Grism~7 spectrum was excluded from the atmospheric-parameter inference and used only as a broadband-continuum check. We performed spectral synthesis with SPECTRUM \citep{GrayCorbally1994} on the Castelli/ATLAS9 grid \citep{Castelli2003}, sampling the joint posterior on $T_{\rm eff}$, $\log g$, and [M/H] by MCMC over 6200--8000~\Ang\ with chromospheric, accretion-sensitive, and telluric regions masked. Informative priors on $T_{\rm eff}$ ($\sigma=400$~K) and $\log g$ ($\sigma=0.3$~dex) from \citet{vanCapelleveen2025} break the $T_{\rm eff}$--$\log g$ degeneracy; [M/H] carries no prior. Veiling is expected to be negligible for any plausible accretion rate consistent with the H$\alpha$ non-detection (Sect.~\ref{sec:halpha}), so we adopt a veiling-free model; an a posteriori check using the upper limit from Sect.~\ref{sec:macc} gives $r_{6000}\lesssim0.01$, confirming the assumption. The fit is unimodal with reduced $\chi_\nu^2=0.72$ (Appendix~\ref{app:posterior}).
Because $R\sim2200$ is below the resolution regime where \texttt{iSpec} is usually applied, and because WISPIT~2 is a PMS star, we validated the pipeline against five Gaia FGK Benchmark Stars and five K-type PMS empirical templates, both degraded to the HFOSC dispersion and resolution. All validation spectra were processed with the identical \texttt{iSpec} configuration used for WISPIT~2. The recovered parameters and systematic-error budget are detailed in Appendix~\ref{app:validation}. These tests support a robust temperature scale at the $\sim$150--180~K level, while showing that $\log g$ and [M/H] are less precise at HFOSC resolution. The PMS templates test the behaviour of young K-star spectra but do not provide independent metallicity standards, so we define the [M/H] systematic from the GBS comparison and treat [M/H] as a low-resolution, model-dependent global estimate.
For WISPIT~2 the analysis yields $T_{\rm eff}=4551\pm150$~K, $\log g=4.32\pm0.18$, and $\mathrm{[M/H]}=-0.17\pm0.16$ (uncertainties dominated by the validation floor). The temperature agrees within the floor with the photometric $4400\pm50$~K of \citet{vanCapelleveen2025}, confirming the K3--K4 classification. Combining $T_{\rm eff}$ with the SED luminosity $L_\star=0.699\pm0.021\,L_\odot$ \citep{vanCapelleveen2025} gives $R_\star=1.34\pm0.09\,R_\odot$, which with the stellar mass $M_\star=1.08\,M_\odot$ \citep{vanCapelleveen2025} implies $\log g\simeq4.22$, in agreement with the spectroscopic value within the systematic floor; we do not bias-correct $\log g$ for the benchmark calibration offset (Appendix~\ref{app:validation}) and adopt $R_\star=1.34\,R_\odot$ throughout. The metallicity, the only parameter without an informative prior, is the first low-resolution spectroscopic estimate for the system; we treat it as a model-dependent global metallicity rather than a precise abundance. Two further consistency checks on these parameters (an independent $T_{\rm eff}$ estimate from the Na\,{\sc i}~D equivalent width and an extinction check from the flux-calibrated slope) are described in Appendix~\ref{app:checks}.
\subsection{H$\alpha$ and the stellar accretion state}
\label{sec:halpha}
\begin{figure}
\centering
\includegraphics[width=\hsize]{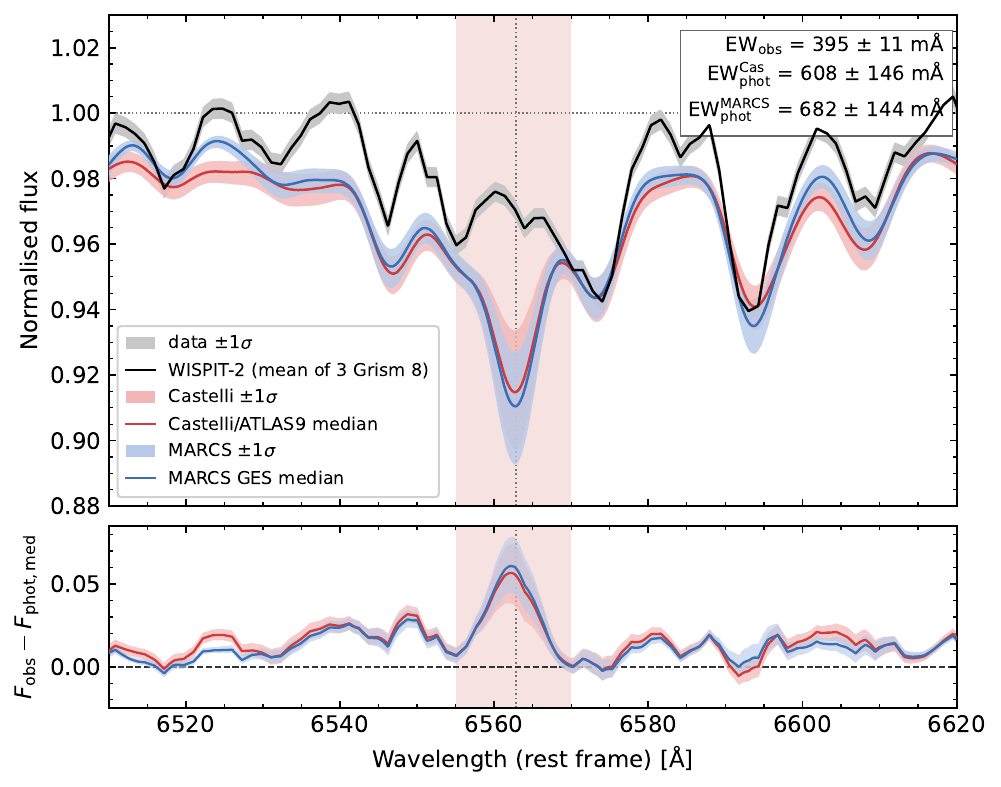}
\caption{Top: normalised H$\alpha$ region of WISPIT~2 (black; mean of three Grism~8 exposures, $\pm 1\sigma$ in grey), with iSpec photospheric syntheses from the Castelli/ATLAS9 (red) and MARCS GES (blue) grids at the posterior-median parameters; coloured bands give the $\pm 1\sigma$ Monte Carlo envelope. The shaded vertical band marks the 6555--6570~\Ang\ EW integration window. Bottom: residuals $F_{\rm obs} - F_{\rm phot}$ for each grid, showing the residual H$\alpha$ filling relative to the photospheric models. The two grids trace each other closely but are systematically offset, giving excess significances of $1.45\sigma$ (Castelli) and $1.98\sigma$ (MARCS).}
\label{fig:halpha}
\end{figure}

H$\alpha$ at 6562.8~\Ang\ remains in net absorption, with no resolved emission core (Fig.~\ref{fig:halpha}). From the three flux-calibrated Grism~8 exposures, combined into an inverse-variance-weighted mean on a common rest-frame grid, we measured the equivalent width by direct integration over 6555--6570~\Ang\ against a local linear pseudo-continuum (6520--6545 and 6580--6605~\Ang\ side-bands), adopting the convention that positive EW denotes net absorption. Per-pixel noise, continuum-window variation ($\pm5$~\Ang), and per-frame scatter were propagated by Monte Carlo. The rest-frame alignment was verified against two clean Fe~{\sc i} lines flanking H$\alpha$. This yields $\mathrm{EW}_{\mathrm{H}\alpha,\,\rm obs} = +0.395 \pm 0.011$~\Ang, with a continuum flux density $F_{\rm cont}(6562.8~\Ang) = (1.375 \pm 0.038) \times 10^{-13}$~erg\,s$^{-1}$\,cm$^{-2}$\,\Ang$^{-1}$ (the $\sim$3\% internal scatter). The absolute flux scale is anchored to the ASAS-SN $g$-band photometry (Appendix~\ref{app:checks}), leaving a residual absolute-calibration uncertainty of $\sim$5\% (the scaling uncertainty), i.e.\ $\sim$0.02~dex in $\log L_{\rm acc}$, sub-dominant to the photospheric-EW systematic.
To estimate the photospheric H$\alpha$ contribution we synthesised the line region with iSpec at the posterior-median parameters (Sect.~\ref{sec:ispec}), using two independent grids (Castelli/ATLAS9 \citep{Castelli2003} and MARCS GES \citep{Gustafsson2008,Heiter2021}) to make the model-atmosphere dependence explicit. For each grid, $N=1000$ Monte Carlo syntheses were drawn from the benchmark systematic floors ($\sigma(T_{\rm eff})=150$~K, $\sigma(\log g)=0.18$, $\sigma([\rm M/H])=0.16$~dex; Sect.~\ref{sec:ispec}), with identical draws for both grids. The EW is dominated by the $T_{\rm eff}$ systematic; $\log g$ and [M/H] contribute negligibly. The two grids give $\mathrm{EW}_{\rm phot}^{\rm Castelli} = 0.608 \pm 0.146$~\Ang\ and $\mathrm{EW}_{\rm phot}^{\rm MARCS} = 0.682 \pm 0.144$~\Ang\ (16th--84th percentile spread), kept separate so the model-atmosphere dependence is explicit. The excess $\mathrm{EW}_{\rm exc} = \mathrm{EW}_{\rm phot} - \mathrm{EW}_{\rm obs}$ is then $+0.213 \pm 0.147$~\Ang\ (Castelli) and $+0.287 \pm 0.144$~\Ang\ (MARCS): positive for both grids but grid-dependent in significance, $1.45\sigma$ and $1.98\sigma$ respectively, the observed EW being $\sim$35--42\% weaker than predicted. We therefore regard the H$\alpha$ residual filling as weak and model-dependent: consistent in sign across both atmosphere grids, but not cleanly separable from chromospheric activity. As we show in Sect.~\ref{sec:macc}, this filling lies $\sim$1.1~dex below the \citet{Manara2013} chromospheric noise expected at this $T_{\rm eff}$, so its accretion-luminosity equivalent is consistent with chromospheric emission alone and yields only an upper limit on the stellar accretion rate.

The inferred $\sim$0.2--0.3~\Ang\ of equivalent filling is far below the $|\mathrm{EW}|>3$~\Ang\ classical-T~Tauri threshold \citep{WhiteBasri2003}, nominally a weak-line T~Tauri star, though the structured disk and directly imaged protoplanets, including the accreting WISPIT~2b, confirm that substantial circumstellar material remains.
\subsection{Stellar accretion rate}
\label{sec:macc}
For reference, and to set an upper limit, we convert the H$\alpha$ residual filling into an accretion luminosity with the \citet{Alcala2017} calibration, propagating all uncertainties (photospheric-EW systematic, observed EW and continuum flux, extinction, distance, and calibration scatter) through a Monte Carlo; the full derivation is given in Appendix~\ref{app:macc}. The dereddened excess gives $\log L_{\rm acc}/L_\odot = -3.54^{+0.26}_{-0.37}$, which lies $\simeq1.1$~dex below the \citet{Manara2013} chromospheric-noise level expected at this $T_{\rm eff}$. The residual H$\alpha$ filling is therefore fully consistent with chromospheric emission, and combined with its low ($1.5$--$2.0\sigma$) significance we treat the H$\alpha$-derived $L_{\rm acc}$ purely as an upper limit, $\dot{M}_{\rm acc} < 3.6\times10^{-11}\,M_\odot\,{\rm yr}^{-1}$ (95\%), far below the canonical $10^{-8}$--$10^{-9}\,M_\odot\,{\rm yr}^{-1}$ CTTS rate \citep{Manara2023}: host accretion is undetectable in this spectrum. The single-epoch measurement may also differ from the time-averaged rate given typical CTTS variability.
\subsection{Lithium and stellar youth}
\label{sec:lithium}
The Li\,{\sc i}~6708~\Ang\ doublet is clearly detected. We measure $\mathrm{EW(Li+Fe)} = 0.344 \pm 0.030$~\Ang\ and infer $\mathrm{EW(Li)} = 0.334 \pm 0.030$~\Ang\ after applying a 10~m\Ang\ blend correction, supporting the independently inferred youth of the star. At HFOSC resolution the corrected EW should still be regarded as an approximate upper limit because of unresolved Fe\,{\sc i}/CN features, so it does not constrain the precise age (Appendix~\ref{app:lithium}).
\section{Discussion}
\label{sec:discussion}
PDS~70, whose two directly imaged protoplanets are both confirmed to be accreting and whose host was characterised with the same \texttt{iSpec} framework \citep{Swastik2021}, is the natural benchmark. WISPIT~2 is hotter (4551 vs.~4152~K) and the two hosts have comparable, mildly sub-solar metallicities ($-0.17$ vs.\ $-0.11$~dex); with $N=2$ we infer no metallicity preference.
WISPIT~2 and PDS~70 compare closely in accretion as well. PDS~70's host accretes at $\sim 10^{-10}\,M_\odot\,{\rm yr}^{-1}$ \citep[$\log\dot{M}_{\rm acc}\approx-10.06$;][]{CampbellWhite2023}, with monitoring spanning $0.6$--$2.2\times10^{-10}\,M_\odot\,{\rm yr}^{-1}$ \citep{Thanathibodee2020}, already over an order of magnitude below the canonical $10^{-8}$--$10^{-9}\,M_\odot\,{\rm yr}^{-1}$ rate for solar-mass T~Tauri stars \citep{Manara2023}. As both hosts have transition disks, the more appropriate comparison is with transition-disk hosts of similar mass; these overlap the CTTS distribution but tend somewhat lower, though not significantly \citep{Manara2023}, and even so the WISPIT~2 limit sits at the extreme low-accretion end. Our upper limit is a factor of $\sim$3 below PDS~70's central rate and below even the lowest epoch of its monitoring \citep{Thanathibodee2020}; with the H$\alpha$ filling $\sim$1.1~dex below the chromospheric noise floor \citep{Manara2013}, it is consistent with pure chromospheric emission and we detect no accretion. Both systems (two embedded gas giants in a structured transition disk around a young roughly solar-mass star) thus appear to show strongly suppressed or undetectable host accretion, on the suppressed-accretion side of a mixed theoretical picture in which gap-opening giants are expected to let gas leak onto the star \citep{Lubow2006,Durmann2015} yet population-synthesis models predict substantial suppression once two or more giants have formed \citep{Manara2019}.
The suppression may indicate that multi-protoplanet disks enter a distinct gas-transport regime of serial filtering, with the outer companion intercepting part of the inflow and the inner companion perturbing the residual, so that overlapping gaps attenuate the channels feeding the star. However, two-planet hydrodynamical simulations give a more conservative picture, in which gas can still diffuse through the gaps efficiently and a second accreting planet need not suppress stellar accretion by more than a factor of a few \citep{BergezCasalou2023}. Whether serial filtering alone explains the suppression thus remains open; additional unresolved planets, inefficient gap-crossing flow, or inner-disk dispersal may be required, and resolved inner-disk gas observations (e.g.\ with ALMA) would provide a direct test.

The host rate can also be compared with the planetary rates. \citet{Close2025} measured $\dot{M}_{\rm 2b}\sim2\times10^{-12}\,M_\odot\,{\rm yr}^{-1}$ for WISPIT~2b, so even our upper limit implies $\dot{M}_{\star}/\dot{M}_{\rm 2b}\lesssim 18$, and the true ratio may be far smaller. This contrasts sharply with ordinary CTTS, where stellar rates exceed protoplanetary ones by three to four orders of magnitude; the rates are non-contemporaneous and differently calibrated, so the ratio is uncertain at the factor-of-several level, but the much smaller contrast is robust and consistent with the protoplanets intercepting much of the inflow.

\section{Conclusions}
\label{sec:conclusions}
From the first optical spectrum of WISPIT~2 we derive $T_{\rm eff} = 4551 \pm 150$~K, $\log g = 4.32 \pm 0.18$, and a low-resolution, model-dependent $\mathrm{[M/H]} = -0.17 \pm 0.16$~dex, validated at HFOSC resolution against Gaia FGK Benchmark Stars and K-type PMS templates \citep{Manara2017}; the gravity and lithium EW are consistent with a pre-main-sequence K~star. H$\alpha$ shows only weak ($1.5$--$2.0\sigma$) residual filling, $\sim$1.1~dex below the chromospheric noise floor \citep{Manara2013} and hence consistent with pure chromospheric emission, so we detect no accretion and set a 95\% upper limit $\dot{M}_{\rm acc} < 3.6\times10^{-11}\,M_\odot\,{\rm yr}^{-1}$, a factor of a few below PDS~70's central rate and below even the lowest epoch of its monitoring \citep{Thanathibodee2020}. WISPIT~2 thus joins PDS~70 as a double-protoplanet host with strongly suppressed or undetectable accretion; with $N=2$ this is suggestive rather than established. A contemporaneous X-shooter/FEROS study independently finds little to no stellar accretion and reports a 4.8-day spectroscopic binary \citep{Burgy2026}; an ALMA measurement of the inner gas reservoir would test whether gas is retained, filtered, or depleted in the inner disk.
\newpage
\section{Data availability}
The reduced HFOSC spectra of WISPIT~2 in FITS format are only available in electronic form at the CDS via anonymous ftp to \url{cdsarc.u-strasbg.fr} (130.79.128.5) or via \url{http://cdsweb.u-strasbg.fr/cgi-bin/qcat?J/A+A/}.
\begin{acknowledgements}
We thank the staff of the Indian Astronomical Observatory, Hanle, and CREST, 
Hosakote, operated by the Indian Institute of Astrophysics, for support during 
the HCT observations, and the ICTS program EXOPLANETS2026 
(ICTS/EXOPLANETS2026/03). S.C.\ is funded by the European Union (ERC, UNVEIL, 
101076613); views and opinions expressed are those of the authors only and do 
not necessarily reflect those of the European Union or the European Research 
Council, and neither can be held responsible for them. S.C.\ acknowledges 
PRIN-MUR 2022YP5ACE. We thank Carlo Manara for insightful comments and Ajay 
Saini and Srikant Bhandari for taking the observations and sending us the data 
promptly. C.S.\ fondly remembers Lalu, a dear companion who now is in the land of unknown and whose presence brightened the months this work was carried out. This work used \texttt{IRAF}, \texttt{iSpec} \citep{BlancoCuaresma2014a}, \texttt{SPECTRUM} \citep{GrayCorbally1994}, \texttt{astropy}, \texttt{numpy}, \texttt{scipy}, and \texttt{emcee}.
\end{acknowledgements}
\bibliographystyle{aa}
\bibliography{refs}
\onecolumn
\makeatletter
\newenvironment{aaminifigure}{\def\@captype{figure}}{}
\newenvironment{aaminitable}{\def\@captype{table}}{}
\makeatother
\begin{appendix}
\footnotesize
\setlength{\textfloatsep}{5pt plus 1pt minus 1pt}
\setlength{\floatsep}{5pt plus 1pt minus 1pt}
\setlength{\intextsep}{5pt plus 1pt minus 1pt}

\section{Combined optical spectrum and atmospheric-parameter fit}
\label{app:posterior}
\begin{figure}[h!]
\centering
\includegraphics[width=0.94\textwidth,height=0.39\textheight,keepaspectratio]{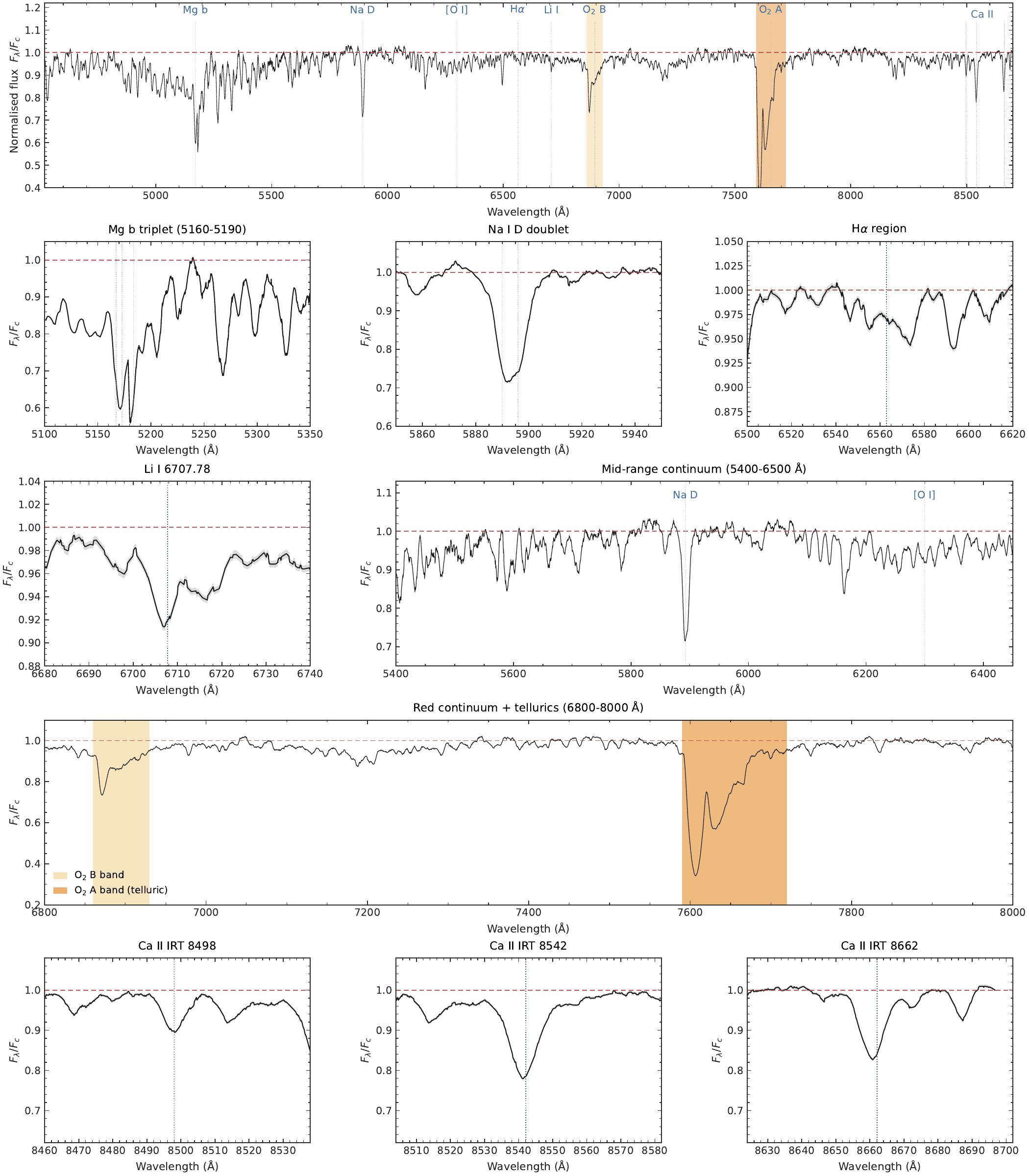}
\caption{Combined, continuum-normalised HFOSC spectrum of WISPIT~2 (Grism~7 + Grism~8, 2026 April 14). Grism~7 is used below 5180~\Ang\ and Grism~8 above it. The panels identify the principal photospheric diagnostics, H$\alpha$, Li\,{\sc i}~6708~\Ang, the Ca\,{\sc ii} infrared triplet, and the O$_2$ telluric bands.}
\label{fig:full_spec}
\end{figure}

\begin{figure}[h!]
\centering
\begin{minipage}[t]{0.30\textwidth}
\centering
\includegraphics[width=0.98\linewidth,height=0.27\textheight,keepaspectratio]{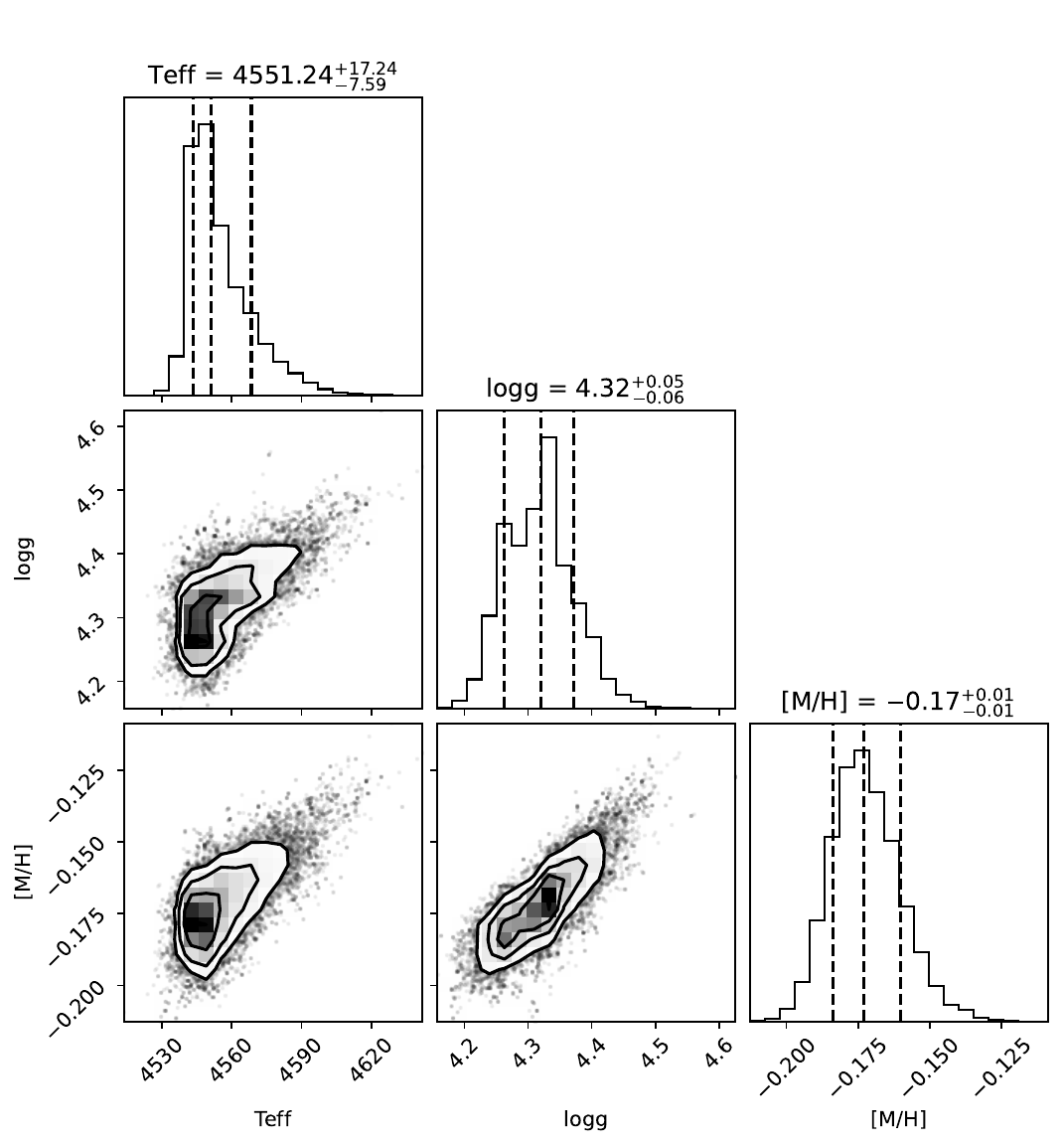}
\caption{MCMC posterior for $T_{\rm eff}$, $\log g$, and [M/H]. Dashed lines mark the 16th, 50th, and 84th percentiles.}
\label{fig:corner}
\end{minipage}\hfill
\begin{minipage}[t]{0.67\textwidth}
\centering
\includegraphics[width=0.99\linewidth,height=0.25\textheight,keepaspectratio]{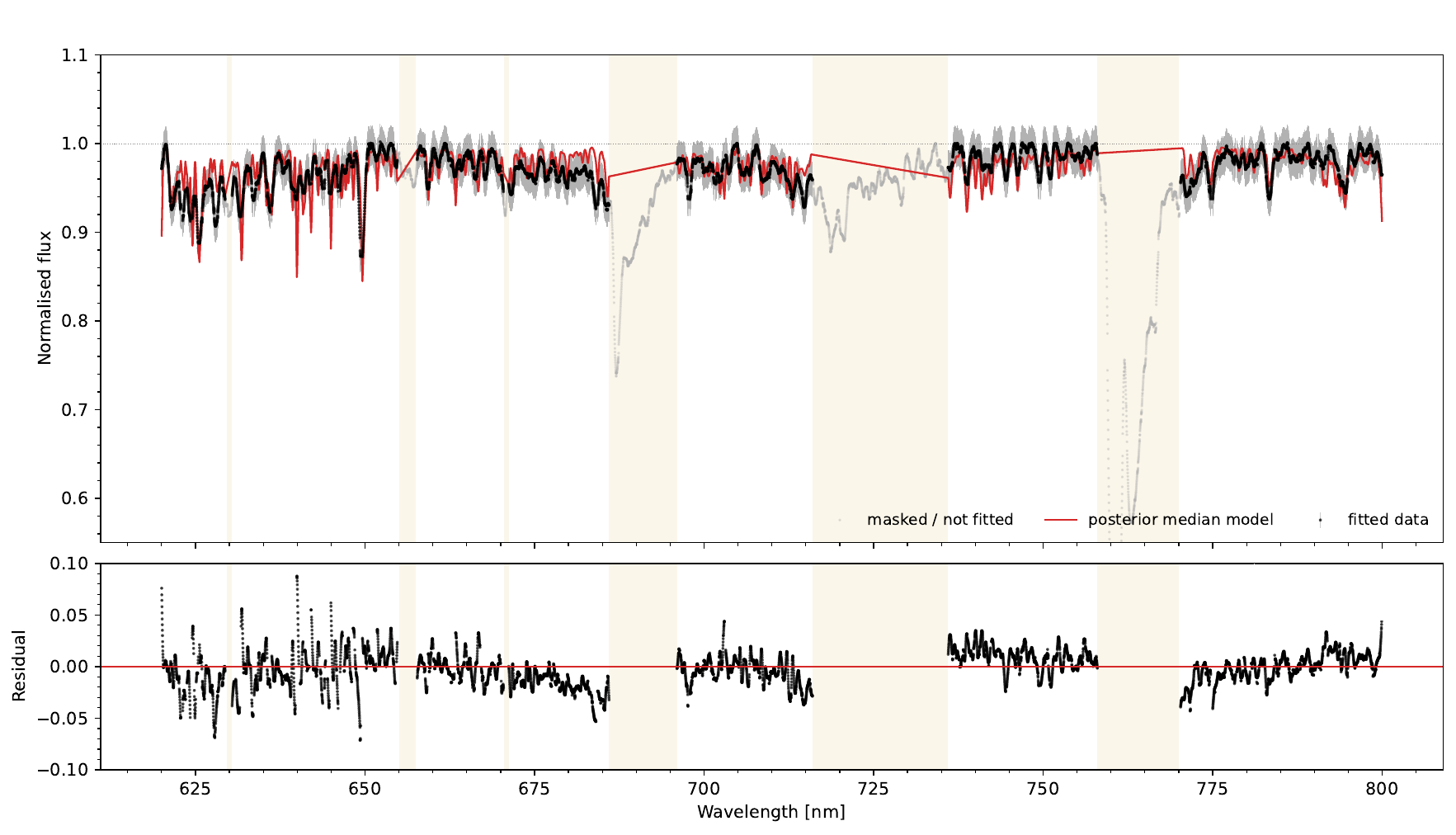}
\caption{Median-combined Grism~8 spectrum (black) and best-fit \texttt{iSpec} model (red) for $T_{\rm eff}=4551$~K, $\log g=4.32$, and $\mathrm{[M/H]}=-0.17$. Grey regions were excluded from the fit. The fit has $\chi_\nu^2=0.72$ and a residual RMS of 1.66\%.}
\label{fig:bestfit}
\end{minipage}
\end{figure}

The continuum-normalised HFOSC spectrum and the atmospheric-parameter fit are summarised in Figs.~\ref{fig:full_spec}--\ref{fig:bestfit}. The posterior is unimodal, with the expected positive correlations of $T_{\rm eff}$ with $\log g$ and [M/H]. Its formal widths do not include the validation-based systematic floors adopted in Sect.~\ref{sec:ispec}.

\clearpage
\footnotesize
\setlength{\baselineskip}{8.2pt}
\setlength{\abovecaptionskip}{2pt}
\setlength{\belowcaptionskip}{1pt}
\section{Benchmark validation of the \texttt{iSpec} pipeline}
\label{app:validation}
We validated the analysis using five Gaia FGK Benchmark Stars (GBS) and five K-type PMS templates from the public Manara et al. libraries \citep{Manara2017}, degraded and analysed with the same settings used for WISPIT~2.

\begin{aaminitable}
\centering
\caption{GBS validation of the \texttt{iSpec} pipeline at HFOSC resolution.}
\label{tab:validation}
\renewcommand{\arraystretch}{0.82}
\resizebox{0.98\textwidth}{!}{
\begin{tabular}{lrrrrrrrrr}
\toprule
Star & $T_{\rm ref}$ & $T_{\rm fit}$ & $\Delta T$ & $g_{\rm ref}$ & $g_{\rm fit}$ & $\Delta g$ & ${\rm [Fe/H]}_{\rm ref}$ & ${\rm [Fe/H]}_{\rm fit}$ & $\Delta{\rm [Fe/H]}$ \\
\midrule
HIP~12114  & 4583 & $4686\pm25$ & $+103$ & 4.52 & $4.73\pm0.04$ & $+0.21$ & $-0.17$ & $-0.19\pm0.07$ & $-0.02$ \\
HIP~104214 & 4398 & $4533\pm21$ & $+135$ & 4.63 & $4.77\pm0.04$ & $+0.14$ & $-0.33$ & $-0.24\pm0.06$ & $+0.09$ \\
HIP~114622 & 4800 & $4677\pm26$ & $-123$ & 4.55 & $4.77\pm0.04$ & $+0.22$ & $+0.04$ & $-0.07\pm0.07$ & $-0.11$ \\
HIP~108870 & 4676 & $4667\pm30$ & $-9$   & 4.59 & $4.77\pm0.04$ & $+0.18$ & $-0.08$ & $-0.15\pm0.08$ & $-0.07$ \\
HIP~73184  & 4657 & $4586\pm28$ & $-71$  & 4.60 & $4.77\pm0.04$ & $+0.17$ & $+0.03$ & $-0.03\pm0.07$ & $-0.06$ \\
\bottomrule
\end{tabular}}
\end{aaminitable}

\begin{aaminitable}
\centering
\caption{PMS-template validation using K-type empirical spectra degraded to HFOSC resolution.}
\label{tab:pms_validation}
\renewcommand{\arraystretch}{0.82}
\resizebox{0.98\textwidth}{!}{
\begin{tabular}{lrrrrrrr}
\toprule
Object & $T_{\rm ref}$ (K) & $T_{\rm fit}$ (K) & $\Delta T$ (K) & $g_{\rm ref}$ & $g_{\rm fit}$ & $\Delta g$ & ${\rm [M/H]}_{\rm fit}$ \\
\midrule
RXJ1538.6$-$3916 & $4522\pm127$ & $4654\pm16$ & $+132$ & $4.21\pm0.13$ & $4.11\pm0.03$ & $-0.10$ & $+0.22\pm0.01$ \\
RXJ1543.1$-$3920 & $4308\pm107$ & $4429\pm15$ & $+121$ & $4.12\pm0.23$ & $4.23\pm0.05$ & $+0.11$ & $+0.17\pm0.01$ \\
RXJ1547.7$-$4018 & $4971\pm75$  & $4793\pm19$ & $-178$ & $4.22\pm0.16$ & $4.07\pm0.05$ & $-0.15$ & $+0.10\pm0.01$ \\
RXJ0438.6$+$1546 & $4915\pm170$ & $4792\pm15$ & $-123$ & $4.12\pm0.30$ & $4.42\pm0.04$ & $+0.30$ & $+0.14\pm0.01$ \\
RXJ1515.8$-$3331 & $4997\pm71$  & $4920\pm11$ & $-77$  & $3.86\pm0.25$ & $4.29\pm0.01$ & $+0.43$ & $+0.04\pm0.01$ \\
\bottomrule
\end{tabular}}
\end{aaminitable}

\noindent GBS reference $T_{\rm eff}$ and $\log g$ values are from \citet{Soubiran2024}, and [Fe/H] from \citet{Jofre2014}; recovered uncertainties are formal posterior widths. The offsets motivate floors of 150~K in $T_{\rm eff}$ and 0.16~dex in [M/H], while the coherent $+0.18\pm0.03$~dex gravity offset is used as a floor rather than a correction. The PMS median absolute offsets are 123~K and 0.15~dex in $T_{\rm eff}$ and $\log g$. Their [M/H] values are only a near-solar consistency check because the templates are not metallicity standards. Veiling was fixed to zero, consistent with $r_{6000}\lesssim0.01$.

\vspace{2pt}
\noindent
\begin{minipage}[t]{0.485\textwidth}
\vspace{0pt}

\section{Flux-calibration, temperature, and extinction checks}
\label{app:checks}
The response-corrected spectra were scaled to the ASAS-SN $g$-band light curve \citep{Shappee2014,Kochanek2017}; the required factor was $2.856\pm0.137$. Equivalent widths are invariant to this scaling, whereas its uncertainty is propagated into the H$\alpha$ luminosity. The Na\,{\sc i}~D equivalent width, $3.7\pm0.2$~\Ang, gives a first-guess $T_{\rm eff}\approx4850$~K from the \citet{Tripicchio1997} relation. Its intrinsic $\sim300$~K scatter and unmodelled gravity and metallicity effects make it consistent with $4551\pm150$~K.

Over 5000--9000~\Ang, the continuum slope formally favours $A_V\sim0.6$, but the optical estimate is imperfect: beyond 8000~\Ang\ the calibration is affected by increased uncertainty in the red response correction. We therefore retain the SED value $A_V=0.171\pm0.050$~mag \citep{vanCapelleveen2025}.

\section{Accretion-luminosity derivation}
\label{app:macc}
A Monte Carlo calculation ($N=10^5$) samples the photospheric-EW systematic, observed EW, continuum flux, 5\% absolute scaling, $A_V$, distance, and the 0.13~dex \citet{Alcala2017} scatter. Negative excesses are retained as zero accretion luminosity. We obtain $F_{\mathrm{H}\alpha,\rm exc}=(3.5\pm1.9)\times10^{-14}$~erg\,s$^{-1}$\,cm$^{-2}$, $\log L_{\mathrm{H}\alpha}/L_\odot=-4.67^{+0.19}_{-0.31}$, and $\log L_{\rm acc}/L_\odot=-3.54^{+0.26}_{-0.37}$. This lies $\simeq1.1$~dex below the \citet{Manara2013} chromospheric-noise level. Interpreting all filling as accretion would give $\dot M_{\rm acc}=1.4^{+1.2}_{-0.8}\times10^{-11}\,M_\odot\,{\rm yr}^{-1}$; because the filling is marginal and chromospheric, we instead adopt the 95\% upper limit stated in Sect.~\ref{sec:macc}.
\end{minipage}\hfill
\begin{minipage}[t]{0.485\textwidth}
\vspace{0pt}
\section{Lithium equivalent width and age comparison}
\label{app:lithium}
The unresolved Li\,{\sc i}~6708~\Ang\ blend was measured in the three Grism~8 exposures. Each spectrum was continuum-normalised over 6680--6698 and 6717--6740~\Ang, and fitted over 6703--6711.5~\Ang\ with $1-f_{\rm norm}(\lambda)=A\exp[-(\lambda-\lambda_0)^2/(2\sigma^2)]$, so $\mathrm{EW}=A\sigma\sqrt{2\pi}$. The coarser Grism~7 exposure was excluded. The measurements are mutually consistent ($\chi^2=1.03$ for two degrees of freedom) and give $\mathrm{EW(Li+Fe)}=0.344\pm0.030$~\Ang. Subtracting a 10~m\Ang\ Fe\,{\sc i}/CN correction yields $\mathrm{EW(Li)}=0.334\pm0.030$~\Ang, an approximate upper limit at this resolution.

\begin{aaminitable}
\centering
\caption{Li\,{\sc i}~6708~\Ang\ equivalent-width measurements.}
\label{tab:li}
\scriptsize
\begin{tabular}{lccc}
\toprule
Spectrum & $t_{\rm exp}$ (s) & EW(Li+Fe) (\Ang) & $\sigma_{\rm EW}$ (\Ang) \\
\midrule
ajd14044 & 600 & 0.317 & 0.046 \\
ajd14045 & 600 & 0.389 & 0.055 \\
ajd14047 & 300 & 0.340 & 0.054 \\
Weighted mean & & 0.344 & 0.030 \\
\bottomrule
\end{tabular}
\end{aaminitable}

The fitted FWHM (4.8--5.7~\Ang) confirms that the doublet and neighbouring blend are unresolved. At $T_{\rm eff}\simeq4551$~K the measured EW lies near the upper Pleiades envelope, but blending and the intrinsic factor-of-two spread among K dwarfs prevent a precise age constraint \citep{Martin1997,Bouvier2018,Somers2017}. Lithium therefore supports youth but does not independently establish the $\sim5$~Myr age, which rests on the SED/isochrone fit and Theia~53 membership \citep{vanCapelleveen2025}.

\begin{aaminifigure}
\centering
\includegraphics[width=0.96\linewidth,height=0.14\textheight,keepaspectratio]{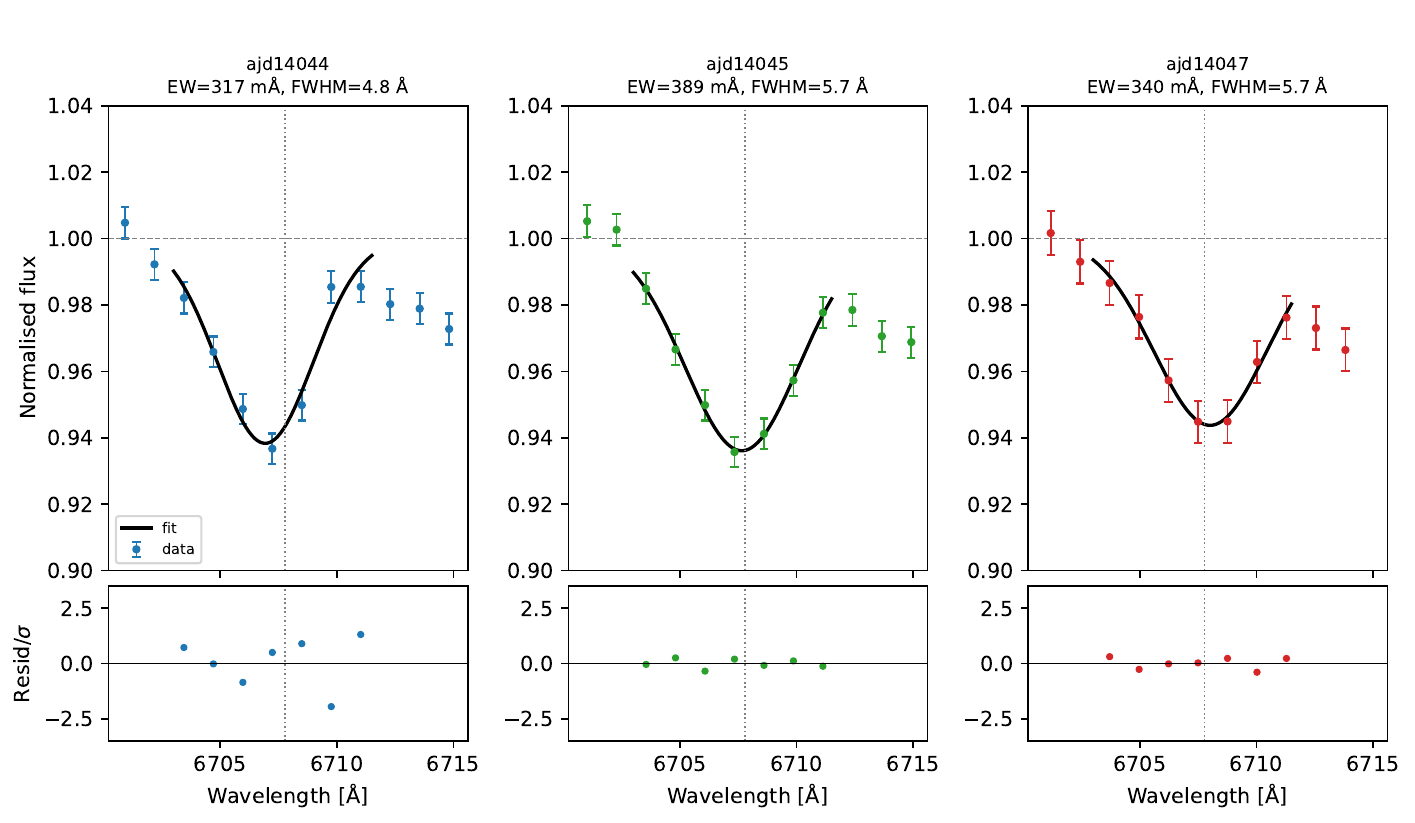}
\caption{Gaussian fits to the Li\,{\sc i}~6708~\Ang\ blend in the three Grism~8 exposures (top) and residuals normalised by the per-pixel noise (bottom). The dotted line marks 6707.78~\Ang.}
\label{fig:lithium_fits}
\end{aaminifigure}
\end{minipage}
\end{appendix}

\end{document}